\documentclass[11pt,twoside,A4]{article}
\usepackage{times,fancyhdr}
\usepackage[dvips]{graphicx}
\usepackage{latexsym}
\usepackage[affil-it]{authblk}
\usepackage{mathptm}
\usepackage{amsmath}
\usepackage{amssymb}
\usepackage{floatrow}
\usepackage{setspace}
\usepackage{hyperref}
\usepackage{graphicx}
\usepackage{wrapfig}
\usepackage[font=small,labelfont=bf]{caption}
\usepackage[top=4cm, bottom=4cm, left=3.5cm, right=3.5cm]{geometry}

\def\beq{\begin{equation}}
\def\eeq{\end{equation}}
\def\beqn{\begin{eqnarray}}
\def\eeqn{\end{eqnarray}}

\begin{document}

\title{Rethinking Superdeterminism}
\author{S.~Hossenfelder$^{1}$, T.~N.~Palmer$^{2}$}
\affil{\small $^{1}$ Frankfurt Institute for Advanced Studies\\
Ruth-Moufang-Str. 1,
D-60438 Frankfurt am Main, Germany
}
\affil{\small $^{2}$ Department of Physics\\
University of Oxford, UK
}

\date{}
\maketitle
\vspace*{-1cm}

\begin{abstract}
Quantum mechanics has irked physicists ever since its conception more than 100 years ago. 
While some of the misgivings, such as it being unintuitive, are merely aesthetic, 
quantum mechanics has one serious shortcoming: it lacks a physical description of the
measurement process. This ``measurement problem''  indicates that quantum 
mechanics is at least an incomplete theory -- good as far as it goes, but missing a piece -- or, more radically, is in need of complete overhaul.

Here we describe an approach which may provide this sought-for completion or replacement: Superdeterminism. A superdeterministic theory is one which violates the assumption of Statistical Independence (that distributions of hidden variables are independent of measurement settings). Intuition suggests that Statistical Independence is an essential ingredient of any theory of science (never mind physics), and for this reason Superdeterminism is typically discarded swiftly in any discussion of quantum foundations. 

The purpose of this paper is to explain why the existing objections to Superdeterminism are based on experience with classical physics and linear systems, but that this experience
misleads us. Superdeterminism is a promising approach not only to solve the measurement problem, but also to understand the apparent 
nonlocality of quantum physics. Most importantly, we will discuss how it may be possible to test this hypothesis in an (almost) model independent way.

\end{abstract}

\section{Introduction}

Until the 1970s, progress in the foundations of physics meant discovering
new phenomena at higher energies, or short distances, respectively. But
progress in high-energy physics has slowed, and may have run its course as 
far as finding solutions to the deep fundamental problems of physics is concerned. 
In the past decades, physicists have not succeeded in solving any of the open problems in the
foundations of their field; indeed it's not even clear we are getting closer to solving them.  
Most notably, we have still not succeeded in synthesizing quantum and gravitational physics, 
or in unraveling the nature of dark matter, problems that have been 
known since the 1930s. 

In this situation it makes sense to go back and look for the path we
did not take, the wrong turn we made early on that led into this seeming dead end. The turn we did not take, we argue here, is resolving the
shortcomings of quantum mechanics. At the least, we need a physical description of the measurement process that accounts for the nonlinearity of quantum measurement. 

The major reason this path has remained largely unexplored is that
under quite general assumptions (defined below)  any theory which solves the measurement 
problem in a form consistent with the principles of relativity, makes it impossible to prepare a
state independently of the detector that will later measure it. If one is not willing to accept
this dependence then -- by virtue of Bell's theorem \cite{Bell} -- one necessarily
has to conclude that a local, deterministic completion of quantum mechanics
is impossible. This, then, requires us to abandon the principles on which general relativity is
based and adds to our difficulty reconciling gravity with the
other interactions.

If one is, by contrast, willing to accept the consequences of realism, reductionism, and determinism, one is led to a theory in which the prepared state of an experiment is never independent of the detector settings. Such theories are known as ``superdeterministic''. We wish to emphasize that superdeterministic theories are {\sl not} interpretations of quantum mechanics. They are, instead, theories more fundamental than quantum mechanics, from which quantum mechanics can be derived. 

Superdeterminism is frequently acknowledged as an experimentally unclosed loophole (see eg \cite{Gallicchio:2013iva}) with which one can explain deterministically the observed violations of Bell's inequality. However, for a variety of reasons, many physicists think Superdeterminism is a non-starter. For example, they argue that Superdeterminism would turn experimenters into mindless zombies, unable to configure their experimental apparatuses freely. 
A similar argument has it that Superdeterminism implies the existence of 
implausible conspiracies between what would otherwise be considered independent processes. Alternatively, it would seemingly lead to causes propagating backwards in time. Above all, so it is claimed, Superdeterminism would fatally undermine the notion of science as an objective pursuit. In short, Superdeterminism is widely considered to be dead in the water. 

The aim of this paper is to re-examine the arguments against Superdeterminism. We will argue that, rather than being an implausible and dismissible loophole, the neglected option of Superdeterminism is the way forward; it's the path we did not take. 

\section{Why?}

The way it is commonly taught, quantum mechanics has two ingredients to its dynamical law: the Schr\"odinger equation and the measurement prescription. The measurement prescription is a projection on a detector eigenstate, followed by re-normalizing the new state to 1. 

This measurement prescription (also sometimes referred to as the ``update'' or ``collapse'' of the wave-function) is not a unitary operation. It preserves probabilities by construction, but it is neither reversible nor linear. The lack of reversibility is not a serious problem: one may interpret irreversibility as a non-physical limit in which one has ignored small but finite residuals that would otherwise make the measurement process reversible. 

Rather, the major problem with the measurement process is that it is nonlinear. If we have a prepared initial state $|\Psi_1 \rangle$ that brings the detector into eigenstate $|\chi_1 \rangle$, and another initial state $|\Psi_2 \rangle$ that brings the detector into eigenstate $|\chi_2 \rangle$, then a linear evolution law would bring a superpostion $(|\Psi_1 \rangle + |\Psi_2 \rangle)/\sqrt{2}$ into a superposition of detector eigenstates -- but this is not what we observe. 

This is problematic because if quantum mechanics was the correct theory to describe the behaviour of elementary particles, then what macroscopic objects like detectors do should be derivable from it. The problem is not merely that we do not know how to make this derivation, it's far worse: the observed nonlinearity of the measurement process tells us that the measurement process is in contradiction with the linear Schr\"odinger equation.

However, there is a simple resolution of this problem of nonlinearity. In its density matrix form, the Schr\"{o}dinger equation is remarkably similar to the classical Liouville equation. So much that, in this form, the Schr\"{o}dinger equation is sometimes referred to as the Schr\"{o}dinger-Liouville or Quantum-Liouville equation, though the historically more correct term is the von Neumann-Dirac equation:
\beqn
\mbox{Liouville equation:} &\quad& \frac{\partial \rho}{\partial t} = \{ H, \rho \}~,\\
\mbox{von Neumann-Dirac equation:} &\quad& i \hbar \frac{\partial \rho}{\partial t} = \left[ H, \rho \right] ~. 
\eeqn 
Here, $H$ is the classical/quantum Hamiltonian of the system, the curly brackets are Poisson brackets, and the square brackets are commutators. $\rho$ is the classical/quantum (probability) density, respectively.

The classical Liouville equation is linear in the probability density due to conservation of probability. But this linearity says nothing whatsoever about whether the dynamics of the underlying system from which the probability density derives is also linear.  Hence, for example, chaotic dynamical systems, despite their nonlinear dynamics, obey the same linear equation for  probability density. To us, this close formal similarity between the two equations strongly suggests that quantum physics, too, is only the linear probabilistic description of an underlying nonlinear deterministic system. 

From this point of view, pursuing an Everettian approach to quantum physics is not the right thing to do, because this idea is founded on the belief that the Schr\"{o}dinger equation is fundamental; that nothing underpins it. Moreover, it does not make sense to just append nonlinear dynamics to the Schr\"{o}dinger equation in situations when state decoherence becomes non-negligible, because  it is not the Schr\"{o}dinger equation itself that needs to become nonlinear. Spontaneous collapse models do not help us either because these are not deterministic\footnote{One may, however, expect spontaneous collapse models to appear as effective descriptions of a non-linear collapse process in suitable limits.}. Pilot-wave theories do, in some sense, solve the measurement problem deterministically. However, pilot-wave formulations of quantum mechanics are based on an explicitly nonlocal ontology, and this nonlocality makes it difficult to reconcile such theories with special relativity and, with that, quantum field theory.

What, then, does it take to describe quantum physics with a deterministic, local theory that is reductionist in the sense that the theory allows us to derive the behaviour of detectors from the behaviour of the theory's primitive elements? The Pusey-Barrett-Rudolph ({\sc PBR}) theorem \cite{Pusey:2011de,Leifer:2014sza} tells us that such a theory must violate the Preparation Independence Postulate, according to which the state space of a composite system can be described as a product state whose factors are independent of each other. Violating Preparation Independence entails that either the systems making up the product state are correlated with each other, or that the composite system cannot be described as a product state to begin with. This lack of independence between the prepared state and the detector is the hallmark of Superdeterminism.

\section{What?}
\label{what}

We define a superdeterministic theory as a Psi-epistemic, deterministic theory that violates Statistical Independence but is local in the sense of respecting Continuity of Action \cite{Wharton}, ie, there is no ``action at a distance'' as Einstein put it. In the remainder of this section we will explain what these words mean.

{\bf 1. Psi-epistemic:} That a theory is Psi-epistemic means that the wave-function in the Schr\"odinger equation (or the density-matrix, respectively) does not itself correspond to an object in the real world, i.e. is not ontic. The Copenhagen interpretation and Neo-Copenhagen interpretations are Psi-epistemic because they postulate the wave-function merely encodes knowledge about the state of the system, rather than itself corresponding to a property of the system. However, a theory may also be Psi-epistemic because the wavefunction is emergent, for example as a statistical representation of a more fundamental theory. The theories we will be dealing with here are Psi-epistemic in the latter sense. 

Needless to say, the wavefunction derived in any such theory should obey the Schr\"odinger equation up to current measurement precision and hence reproduce the so-far tested predictions of quantum mechanics. But of course the point of seeking a theory from which to derive quantum mechanics is not to reproduce quantum mechanics, but to make predictions beyond that.

{\bf 2. Deterministic:} By deterministic we will mean that the dynamical law of the theory uniquely maps states at time $t$ to states at time $t'$ for any $t$ and $t'$. This map, then, can be inverted. 

Since the theory we look for should be deterministic and the wavefunction derives from it, we are dealing with a so-called hidden-variable theory. We can ask what exactly are these hidden variables, which in the following are collectively represented by the symbol $\lambda$. The answer depends on the specific model one is dealing with, but loosely speaking $\lambda$ contains all the information that is required to determine the measurement outcome (except the ``not hidden'' variables that are the state preparation). In this picture, quantum mechanics is not deterministic simply because we do not know $\lambda$.

It is important to realise that these hidden variables are not necessarily properties intrinsic to or localised within the particle that one measures; they merely have to determine the outcome of the measurement. To see the distinction, consider the following example. You are standing in a newborn ward in a hospital and look at a room full of screaming infants. On your mind are two questions: What's their blood type? and Will they ever climb Mount Everest? In a deterministic theory, answers to both questions are encoded in the state of the universe at the present time, but they are very different in terms of information availability. A baby's blood type is encoded locally within the baby. But the information about whether a baby will go on to climb Mount Everest is distributed over much of the hypersurface of the moment the baby is born. It is not, in any meaningful sense, an intrinsic property of the baby. This example also illustrates that just because a theory is deterministic, its time evolution is not necessarily predictable. 

{\bf 3. Violation of Statistical Independence:} The most distinctive feature of superdeterministic theories is that they violate Statistical Independence. As it is typically expressed,  this means that the probability distribution of the hidden variables, $\rho(\lambda)$, is not independent of the detector settings. If we denote the settings of two detectors in a Bell experiment as {\bf a} and {\bf b}, we can write this as
\beqn
\rho (\lambda |{\bf a}, {\bf b}) \neq \rho (\lambda) ~.
\eeqn
For example, in the {\sc CHSH} version of Bell's Theorem \cite{CHSH}, $\bf a$ and $\bf b$ each take one of two discrete orientations, which we can represent here as 0 or 1. To derive Bell's inequality, one assumes $\rho (\lambda | {\bf a}, {\bf b}) = \rho (\lambda)$, a requirement that is also often referred as ``Free Choice''. (This terminology is profoundly misleading as we will discuss in Section \ref{FreeWill}.)

While it is straightforward to write down Statistical (In)dependence as a mathematical requirement, the physical interpretation of this assumption less clear.
One may be tempted to read the probability encoded by $\rho$ as a frequency of occurrence for different combinations of $(\lambda, {\bf a}, {\bf b})$ that happen in the real world. However, without further information about the theory we are dealing with, we do not know whether any particular combination ever occurs in the real world. E.g., in the case that a pair of entangled particles is labelled by a unique $\lambda$, for any value of $\lambda$ only one pair of values for $\bf a$ and $\bf b$ would actually be realised in the real world. 

At the very least, whether these alternative combinations of hidden variables and detector settings ever exist depends both on the state space of the theory and on whether dynamical evolution is ergodic on this state space. It is easy to think of cases where dynamical evolution is not ergodic with respect to the Lebesgue measure on state space. Take for example a classical, nonlinear system, like the iconic Lorenz model \cite{Lorenz}. Here, the asymptotic time evolution is constrained to an attractor with fractal measure, of a dimension lower than the full state space. For initial conditions on the attractor, large parts of state space are never realized.

Neither can we interpret $\rho$ as a probability in the Bayesian sense\footnote{As is the idea behind QBism \cite{qbism}.}, for then it would encode the knowledge of agents and thereby require us to first define what ``knowledge'' and ``agents'' are. This interpretation, therefore, would bring back the very difficulty we set out to remove, namely that a fundamental theory for the constituents of observers should allow us to derive macroscopic concepts.

We should not, therefore, interpret Statistical Independence as a statement about properties of the real world, but understand it as a mathematical assumption of the model with which we are dealing. This point was made, implicitly at least, by Bell himself \cite{Bell}:
\begin{quote}
``I would insist here on the distinction between analyzing various physical theories, on the one hand, and philosophising about the unique real world on the other hand. In this matter of causality it is a great inconvenience that the real world is given to us once only. We cannot know  what would have happened if something had been different. We cannot repeat an experiment changing just one variable; the hands of the clock will have moved, and the moons of Jupiter. Physical theories are more amenable in this respect. We can \emph{calculate} the consequences of changing free elements in a theory, be they only initial conditions, and so can explore the causal structure of the theory. I insist that [Bell's Theorem] is primarily an analysis of certain kinds of theory.'' (emphasis original)
\end{quote}

In summary, Statistical Independence is not something that can be directly tested by observation or by experiment because it implicitly draws on counterfactual situations, mathematical possibilities that we do not observe and that, depending on one's model or theory, may or may not exist. 

\begin{wrapfigure}{r}{0.45\textwidth}
  \begin{center}
    \includegraphics[width=0.88\textwidth]{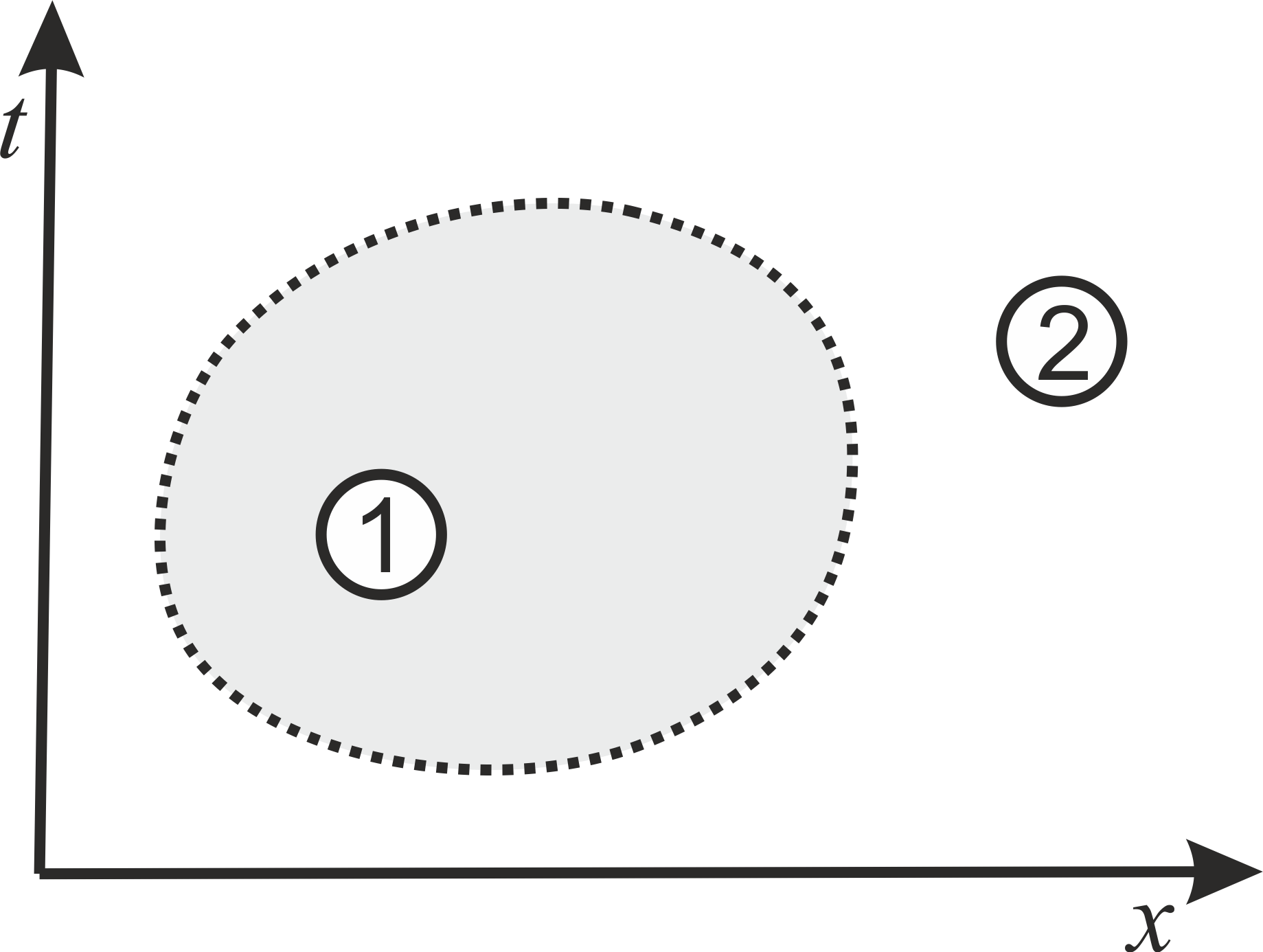}
  \end{center}
  \caption{Continuity of Action. If information from event 1 can influence event 2, then this information must affect any closed three surface around 1. \label{fig2} }

\end{wrapfigure}

{\bf 4. Locality:} Finally, we will assume that the superdeterministic theory respects Continuity of Action (for extensive discussion of the term, see \cite{Wharton}). Continuity of Action (hereafter  CoA) means that to transfer information from one space-time region to another, disjoint, region, the same information has to also be present on any closed (3-dimensional) surface surrounding the first region (see Fig. \ref{fig2}). Information, here, refers to quantities that are locally measurable. We make this assumption because both general relativity and quantum field theories respect this criterion. 
%
%

As laid out in \cite{Wharton}, the definition of locality by CoA is not as strong as the locality assumptions entering Bell's theorem. Besides Statistical Independence, the assumptions for Bell's theorem are 
\begin{enumerate}
\item {\bf Output Independence}\\
This assumption states that the measurement outcome is determined by hidden variables, $\lambda$, and that the hidden variables are the origin of statistical correlations between distant measurement outcomes.  Formally it says that  the distribution for the measurement outcomes $x_{\bf a}$ of detector {\bf a} does not depend on the distribution of outcomes $x_{\bf b}$ at detector {\bf b} and vice versa, ie $\rho_{\bf ab} (x_{\bf a}, x_{\bf b}| {\bf a},{\bf b},\lambda) = \rho_{\bf a} (x_{\bf a}| {\bf a}, {\bf b}, \lambda) 
\rho_{\bf b} (x_{\rm b} | {\bf a}, {\bf b},\lambda)$.

\item {\bf Parameter Independence}\\
Parameter independence  says that the probability distribution of measurement outcomes at one detector does not depend on the settings of the other detector, ie they can be written as $\rho_{\bf a} (x_{\bf a}| {\bf a}, {\bf b}, \lambda) = \rho_{\bf a} (x_{\bf a}| {\bf a}, \lambda)$ and $\rho_{\bf b} (x_{\bf a}| {\bf a}, {\bf b}, \lambda) = \rho_{\bf b} (x_{\bf b}| {\bf b}, \lambda)$. 
\end{enumerate}
These two assumptions together are also known as ``Factorization''. The observed violations of Bell's inequality then imply that at least one of the three assumptions necessary to derive the inequality must be violated. Quantum mechanics respects Statistical Independence and Parameter Independence but violates Outcome Independence. Superdeterminism violates Statistical Independence. Bell-type tests cannot tell us which of the two options is correct. 

All three assumptions of Bell's theorem -- Statistical Independence, Output Independence, and Parameter Independence -- are sometimes collectively called ``local realism'' or ``Bell locality''. However, Bell's local realism has little to do with how the term ``locality'' is used in general relativity or quantum field theory, which is better captured by CoA.  It has therefore been proposed that Bell locality should better be called Bell separability \cite{Hall}. However, that terminology did not catch on.


The issue of whether Factorization is a suitable way to encode locality and causality is similar to the issue with interpreting Statistical Independence: It draws on alternative versions of reality that may not ever actually occur. Factorization requires us to ask what the outcome of a measurement at one place would have been given another measurement elsewhere (Outcome Independence) or what the setting of one detector would have been had the other detector's setting been different (Parameter Independence). These are virtual changes, expressed as state-space perturbations. The changes do therefore not necessarily refer to real events happening in space-time. By contrast, Continuity of Action, is a statement about processes that do happen in space-time: its definition does not necessarily invoke counterfactual worlds. 

To make this point in a less mathematical way, imagine Newton clapping his hands and hearing the sound reflected from the walls of his College quad (allowing Newton to estimate the speed of sound). He might have concluded that the reflected sound was caused by his clap either because:
\begin{itemize}
\item if he had not clapped, he would not have heard the sound;
\item the clapping led to the excitation of acoustic waves in the air, which reflected off the wall and propagated back to vibrate Newton's ear drums, sending electrical signals to his brain.
\end{itemize}
The first definition of causality here depends on the existence of counterfactual worlds and 
with that on the mathematical structure of one's theory of physics. It makes a statement that is impossible to experimentally test.

The second of these definitions, in contrast, identifies a causal connection between the clap and the cognitive recognition: there is no closed region of space-time surrounding the clap that is not affected by either the acoustic wave or the electrical signal.  It is a statement about what actually happens. 

This tension between space-time based notions of causality and the assumptions of Bell's theorem was recently highlighted by a new variant of Bell's theorem for temporal order \cite{Zych:2017tau}. While the authors suggest their theorem shows that a quantum theory of gravity (under certain assumptions listed in the paper) must lack causal order, the theorem equivalently says that if one wants the weak-field limit of quantum gravity to have a well-defined causal order, then Statistical Independence must be violated.

In summary, by relying on Continuity of Action instead of Factorization we avoid having to make statements about non-observable versions of our world. 

\subsection{Retrocausality and Future Input Dependence}

The violation of Statistical Independence, which superdeterministic theories display, implies a correlation between the detector and the prepared state (as defined by $\lambda$), typically in space-like separated regions. These regions, however, are contained in a common causal diamond, so there are events in the past which have both regions in their causal future, and there are events in the future which have both regions in the past. 

The possibility that both detector and prepared state are correlated because of a common event in the past is commonly referred to as the ``locality loophole'' (to Bell's theorem). One can try to address it (to some extent) by choosing the detector settings using events in the far past. An extreme version of this has been presented in \cite{Handsteiner:2016ulx} where light from distant quasars was used to select detector settings. We have more to say about what this experiment does and does not prove in Section \ref{photon}.

The possibility that detector and prepared state are correlated because of an event in the future is often referred to as ``retrocausality'' or sometimes as ``teleology.'' Both of these terms are misleading. The word ``retrocausality'' suggests information traveling backward in time, but no superdeterministic model has such a feature. In fact, it is not even clear what this would mean in a deterministic theory. Unless one explicitly introduces an arrow of time (eg from entropy increase), in a deterministic theory, the future ``causes'' the past the same way the present ``causes'' the future. The word ``teleology'' is commonly used to mean that a process happens to fulfill a certain purpose which suffices to explain the process. It is misleading here because no one claims that Superdeterminism is explanatory just because it gives rise to what we observe; this would be utterly non-scientific. A superdeterministic theory should of course give rise to predictions that are not simply axioms or postulates

For this reason, it was suggested in \cite{Wharton} to use the more scientific expression ``Future Input Dependence.'' This term highlights that to make predictions with a superdeterministic model one may use input on a spacelike hypersurface to the future of system preparation, instead of using input at the time of preparation. This is possible simply because these two slices are connected by a determinstic law. Relying on ``future input'' may sound odd, but it merely generalizes the tautologically true fact that to make a prediction for a measurement at a future time, one assumes that one makes a measurement at a future time. That is, we use ``future input'' every time we make a measurement prediction. It is just that this input does not normally explicitly enter the calculation. 

We wish to emphasize that Future Input Dependence is in the first place an operational property of a model. It concerns the kind of information that one needs to make a prediction. In contrast with the way we are used to dealing with models, Future Input Dependence allows the possibility that this information may arise from a boundary condition on a future hypersurface. Of course one does not actually know the future. However, drawing on future input allows one to make conditional statements, for example of the type ``if a measurement of observable $O$ takes place, then...''. Here, the future input would be that observable $O$ will be measured in the first place. (Of course in that case one no longer predicts the measurement of $O$ itself.)

Now, in a deterministic theory, one can in principle formulate such future boundary conditions in terms of constraints on an earlier state. But the constraint on the earlier state may be operationally useless. That is to say, whether or not one allows Future Input Dependence can make the difference between whether or not a model has explanatory power. (See Section
\ref{cons} for more on that.) 

Future Input Dependence is related to Superdeterminism because constraints on a future hypersurface will in general enforce correlations on earlier hypersurfaces; ie future input dependent theories generically violate Statistical Independence.

In summary, superdeterministic models are not necessarily either retrocausal or teleological (and indeed we are not aware of any model that exhibits one of these properties). But superdeterministic models may rely on future input to make conditional predictions.

\subsection{Disambiguation}

The reader is warned that the word ``Superdeterminism'' has been used with slightly different meaning elsewhere. In all these meanings, Statistical Independence is violated and the corresponding theory should prohibit action at a distance. But some authors  \cite{Wharton,Price} distinguish Superdeterminism from retrocausality (or Future Input Dependence, respectively). Further, not everyone also assumes that a superdeterministic theory is deterministic in the first place. We here assume it is, because this was historically the motivation to consider this option, and because if the theory was not deterministic there would not be much point in considering this option.

\section{Common Objections to Superdeterminism}

In this section we will address some commonly raised objections to Superdeterminism found in various places in the literature and online discussions. 

\subsection{Free Will and Free Choice}
\label{FreeWill}

The Statistical Independence assumption is often referred to as ``Free Choice,'' because it can be interpreted to imply that the experimenter is free to choose the measurement setting independently of the value of the hidden variables. This has had the effect of anthropomorphising what is merely a mathematical assumption of a scientific hypothesis. Let us
therefore have a look at the relation between Statistical Independence and the physical processes that underlie free choice, or free will more generally. 

Ever since Hume \cite{Hume}, the notion of free will has been defined in two different ways:
\begin{itemize}
\item as an ability to have done otherwise;
\item as an absence of constraints preventing one from doing what one wishes to do.
\end{itemize}
As in our previous discussion of causality, these definitions are profoundly different in terms of physical interpretation. An ability to have done otherwise presumes that a hypothetical world where one did do otherwise is a physically meaningful concept. That is to say, the scientific meaningfulness of the notion of `an ability to have done otherwise' depends on the extent to which one's theory of physics supports the notion of counterfactual worlds: as discussed below, theories may vary as to this extent.

The second definition, by contrast, does not depend on the existence of counterfactual worlds. It is defined entirely in terms of events or processes occurring in spacetime. For example, what one ``wishes to do" could be defined in terms of a utility function which the brain attempts to optimise in coming to what we can call a ``choice'' or ``decision''. This second definition is often referred to as the compatibilist definition of free will.

Statistical Independence relies on the first of these definitions of free will because (as discussed above) it draws on the notion of counterfactual worlds. The absence of Statistical Independence does not, however, violate the notion of free will as given by the second definition. We do not usually worry about this distinction because in the theories that we are used to dealing with, counterfactuals typically lie in the state space of the theory. But the distinction becomes relevant for superdeterministic theories which may have constraints on state-space that rule out certain counterfactuals (because otherwise it would imply internal inconsistency). In some superdeterministic models there are just no counterfactuals in state space (for an example, see Section \ref{thooft}), in some cases counterfactuals are partially constrained (see Section \ref{IST}), in others, large parts of state-space have an almost zero measure (\ref{path}).

One may debate whether it makes sense to speak of free will even in the second case since a deterministic theory implies that the outcome of any action or decision was in principle fixed at the beginning of the universe. But even adding a random element (as in quantum mechanics) does not allow human beings to choose one of several future options, because in this case the only ambiguities about the future evolution (in the measurement process) are entirely unaffected by anything to do with human thought. Clearly, the laws of nature are a constraint that \emph{can} prevent us from doing what we want to do. To have free will, therefore, requires one to use the compatibilist notion of free will, even if one takes quantum mechanics in its present form as fundamental. Free will is then merely a reflection of the fact that no one can tell in advance what decisions we will make. 

But this issue with finding a notion of free will that is compatible with deterministic laws (or even partly random laws) is not specific to Superdeterminism. It is therefore not an argument that can be raised against Superdeterminism. Literally all existing scientific theories suffer from this conundrum. Besides, it is not good scientific practice to discard a scientific hypothesis simply because one does not like its philosophical implications.

Let us look at a simple example to illustrate why one should not fret about the inability of the experimenter to prepare a state independently of the detector. Suppose you have two fermions. The Pauli exclusion principle tells us that it is not possible to put these two particles into identical states. One could now complain that this violates the experimenter's free will, but that would be silly. The Pauli exclusion principle is a law of nature; it's just how the world is. Violations of Statistical Independence, likewise, merely tell us what states can exist according to the laws of nature. And the laws of nature, of course, constrain what we can possibly do. 

In summary, raising the issue of free will in the context of Superdeterminism is a red herring.  Superdeterminism does not make it any more or less difficult to reconcile our intuitive notion of free will with the laws of nature than is the case for the laws we have been dealing with for hundreds of years already.

\subsection{The Conspiracy Argument}
\label{cons}

This argument has been made in a variety of ways, oftentimes polemically. Its most rigorous version can be summarised as follows. In any deterministic theory one can take a measurement outcome and, by using the law of time-evolution, calculate the initial state that would have given rise to this outcome. One can then postulate that since this initial state gave rise to the observation, we have somehow ``explained'' the observation. If one were to accept this as a valid argument, this would seemingly invalidate the science method in general. For then, whenever we observe any kind of regularity -- say a correlation between X-ray exposure and cancer -- we could say it can be explained simply because the initial state happened to be what it was.

The more polemic version of this is that in a superdeterministic theory, the universe must have been ``just so'' in order that the decisions of experimenters happen to reproduce the predictions of quantum mechanics every single time. Here, the term ``just so'' is invoked to emphasise that this seems intuitively extremely unlikely and therefore Superdeterminism relies on an implausible ``conspiracy'' of initial conditions that does not actually explain anything.

To address this objection, let us first define ``scientific explanation'' concretely to mean that the theory allows one to calculate measurement outcomes in a way that is computationally simpler than just collecting the data. This notion of  ``scientific explanation'' may be too maths-centric to carry over to other disciplines, but will serve well for physics. The criticism levelled at Superdeterminism is, then, that if one were to accept explaining an observation merely by pointing out that an initial state and a deterministic law exists, then one would have to put all the information about the observation already in the initial state, meaning the theory is not capable of providing a scientific explanation in the above defined sense.

One problem with this argument is that just by knowing a theory violates Statistical Independence one cannot tell anything about its explanatory power. For this one needs to study a concrete model. One needs to know how much information one has to put into the initial state and the evolution law to find out whether a theory is or is not predictive. 

Let us look at a specific example from Bell himself \cite{Bell:1987hh}. Bell himself realised that free will was a red herring (see section \ref{FreeWill}) and for that reason his arguments against Superdeterminism are framed in a completely deterministic setting. He imagines that the measurement setting ($\bf a =0$ or $\bf a=1$) is determined by a pseudo-random number generator whose output is exquisitely sensitive to its input $x$ in the sense that the setting depends on the parity of the millionth digit in the decimal expansion of $x$. Bell concludes that whilst the millionth digit indeed determines the measurement settings, it seems implausible to imagine that it systematically influences, or is systematically influenced by, anything else in the universe -- the particle's hidden variables in particular. 

Of course ``it seems implausible'' is not a convincing argument, as Bell himself conceded, writing  \cite{Bell:1987hh}:
\begin{quote}
Of course it might be that these reasonable ideas about physical 
randomizers are just wrong - for the purpose at hand. A theory may 
appear in which such conspiracies inevitably occur, and these 
conspiracies may then seem more digestible than the non-localities of 
other theories. When that theory is announced I will not refuse to 
listen, either on methodological or other grounds. 
\end{quote}
But Bell's intuition rests on the assumption that because worlds which differ only in the millionth digits of the random numbers are very similar to each other, they are necessarily ``close'' to each other. Such statements therefore implicitly depend on the notion of a distance in state-space. We intuitively tend to assume distance measures are Euclidean, but this does not need to be so in state-space. 

Such conspiracy arguments are also often phrased as worries about the need to ``fine-tune'' -- i.e., choose very precisely -- the initial conditions. (See \cite{finetuning} for a quantifiable definition.) The reference to fine-tuning, however, is misleading. There need be nothing {\sl a priori} unscientific about a fine-tuned theory \cite{Hossenfelder:2018ikr}. A fine-tuned theory {\sl may} be unscientific if one needs to put a lot of information into the initial condition thereby losing explanatory power. But this does not necessarily have to be the case. In fact, according to currently accepted terminology both the standard model of particle physics and the concordance model of cosmology are ``fine-tuned'' despite arguably being scientifically useful. 

One way to avoid that fine-tuning leads to a lack of explanatory power is to find a measure that can be defined in simple terms and that explains which states are ``close'' to each other and/or which are distant and have measure zero, i.e., are just forbidden. (See \cite{rfinetuning} for an example of how this negates the problem of \cite{finetuning}.) 

Bell's and similar examples that rest on arguments from fine-tuning (or sensitivity, or conspiracy) all implicitly assume that there is no simple way to mathematically express the allowed (or likely) initial states that give rise to the predictions of quantum mechanics. See also Section \ref{disc} for further discussion on the notion of ``closeness'' in state-space and Section \ref{IST} for an example of a theory where intuitive Euclidean ideas about closeness of worlds fail.

But assuming that something is impossible does not prove that it is impossible. Indeed, it is provable that it is unprovable to show such theories are unscientific because that is just a rephrasement of Chaitin's imcompletness theorem \cite{Chaitin}. This theorem, in a nutshell, says that one can never tell that there is no way to further reduce the complexity of a string (of numbers). If we interpret the string as encoding the initial condition, this tells us that we cannot ever know that there is not some way to write down an initial state in a simpler way. 

This is not to say that we can rest by concluding that we will never know that a useless theory cannot be made more useful. Of course, to be considered scientifically viable (not to mention interesting) a superdeterministic theory must actually have an explanatory formulation. We merely want to emphasize that the question whether the theory is scientific cannot be decided merely by pointing out that it violates Statistical Independence.

\subsection{The Cosmic Bell Test and the BIG Bell Test}
\label{photon}

In the Cosmic Bell Test \cite{Handsteiner:2016ulx}, measurement settings are determined by the precise wavelength of light from distant quasars, sources which were causally disconnected at the time the photons were emitted. It is without doubt a remarkable experimental feat, but this (and similar) experiments do not -- cannot -- rule out Superdeterminism; they merely rule out that the observed correlations in Bell-type tests were locally caused by events in the distant past. It is, however, clear from the derivation of Bell's theorem that violations of Bell's inequality cannot tell us whether Statistical Independence was violated. Violations of Bell's inequality can only tell us that at least one of the assumptions of the theorem was violated.

The belief that such tests tell us something about (the implausibility of) Superdeterminism goes back, once again, to the idea that a state which is intuitively ``close'' to the one realized in nature  (eg, the wavelength of the light from the distant quasar was a little different, all else equal) is allowed by the laws of nature and likely to happen. However, in a superdeterministic theory what seems intuitively like a small change will generically result in an extremely unlikely state; that's the whole point. For example, in a superdeterministic theory, a physically possible counterfactual state in which the wave-length of the photon was slightly different may also require changes elsewhere on the past hypersurface, thereby resulting in the experimenter's decision to not use the quasar's light to begin with. 

Similar considerations apply to all other Bell-type tests \cite{Leung:2017ndn,Rauch:2018rvx} that attempt to close the freedom-of-choice loophole, like the {\sc BIG} Bell test \cite{bigbell}. This experiment used input from 100,000 human participants playing a video game to choose detector settings, thereby purportedly ``closing the ‘freedom-of-choice loophole’ (the possibility that the setting choices are influenced by ‘hidden variables’ to correlate with the particle properties.)'' Needless to say, the experiment shows nothing of that type; one cannot prove freedom of choice by assuming freedom of choice. 

In fact, the details of these experiments do not matter all that much. One merely has to note that measuring violations of Bell's inequality, no matter how entertaining the experimental setup, cannot tell us which of the assumptions to the theorem were violated.

\subsection{The Tobacco Company Syndrome}

Finally, let us turn to the claim that the assumption of Statistical Independence in Bell's theorem can be justified by what it would imply in classical physics. This argument is frequently put forward with the example of using a randomized trial to demonstrate that lung cancer is linked to smoking. If one were to allow violations of Statistical Independence in Bell-type experiments, so the argument goes, tobacco companies could claim that any observed correlation between lung cancer and smoking was due to a correlation between the randomization and the measured variable (ie, the incidence of cancer). We do not know where this argument originated, but here are two examples: 
\begin{quote}
``It is like a shill for the tobacco industry first saying that smoking does not cause cancer, rather there is a common cause that both predisposes one to want to smoke and also predisposes one to get cancer (this is already pretty desperate), but then when confronted with randomized experiments on mice, where the mice did not choose whether or not to smoke, going on to say that the coin flips (or whatever) somehow always put the mice already disposed to get lung cancer into the experimental group and those not disposed into the control. This is completely and totally unscientific, and it is an embarrassment that any scientists would take such a claim seriously.''  $\sim$ Tim Maudlin \cite{Tobacco2}
\end{quote}

\begin{quote}
``I think this assumption [of Statistical Independence] is necessary to even do science, because if it were not possible to probe a physical system independently of its state, we couldn't hope to be able to learn what its actual state is. It would be like trying to find a correlation between smoking and cancer when your sample of patients is chosen by a tobacco company.
'' 
$\sim$ Mateus Ara\'ujo \cite{Tobacco1}
\end{quote}

One mistake in the argument against Superdetermism is the claim that theories without the assumption of Statistical Independence are unscientific because they are necessarily void of explanatory power. We already addressed this in Subsection \ref{cons}. However, the tobacco company analogy brings in a second mistake, which is the idea that we can infer from the observation that Statistical Independence is useful to understand the properties of classical systems, that it must also hold for quantum systems. This inference is clearly unjustified; the whole reason we are having this discussion is that classical physics is {\sl not} sufficient to describe the systems we are considering.

We have already mentioned an example of how our classical intuition can fail in the quantum case. This example provides a further illustration. For the tobacco trial, we have no reason to think that multiple realizations of the randomization are impossible. For example, two different randomly drawn sub-ensembles of volunteers (say the first drawn in January, the second in February) can be expected to be statistically equivalent. It is only when our theoretical interpretation of an experiment requires us to consider counterfactual worlds, that differences between classical and quantum theories can emerge.

It is further important to note that the assumption of Statistical Independence does not require ensembles of different, actually occurring experiments (as opposed to virtual experiments that only appear in the mathematics) to have different statistical properties. Consider two ensembles of quantum particles, each measured with different measurement settings (say the first in January, the second in February). Since there is no reference to counterfactuals in this description, we cannot infer that the statistical properties of the hidden variables are any different in the January and February ensembles, even in a theory of quantum physics which violates Statistical Independence. At this level, therefore, there is no difference between the quantum and classical example. In a theory that violates Statistical  Independence, one merely cannot infer that if February's ensemble of particles had been measured with January's measurement settings, the result would have been statistically identical. By contrast, if February's volunteers had been tested in January, we would, by classical theory, have expected statistically identical results. In this sense, the tobacco trial analogy is misleading because it raises the impression that the assumption of Statistical Independence is more outlandish than it really is.

\section{How}

The history of Superdeterminism is quickly told because the topic never received much attention. Already Bell realized that if one observes violations of his inequality, this does not rule out local\footnote{In the sense of Continuity of Action, see Section \ref{what}.}, deterministic hidden variable models because Statistical Independence may be violated \cite{Bell:1987hh}. It was later shown by Brans that if Statistical Independence is violated, any Bell-nonlocal distribution of measurement outcomes can be obtained in EPR-type experiments\cite{brans}. It has since been repeatedly demonstrated that it requires only minute violations of Statistical Independence to reproduce the predictions of quantum mechanics locally and deterministically \cite{Gisin,Hall2,Friedman:2018byq}. 

The scientific literature contains a number of toy models that provide explicit examples for how such violations of Statistical Independence can reproduce quantum mechanics \cite{brans,degorre,Hall3} which have been reviewed in \cite{Hall} (Section 4.2). Toy models which violate Statistical Independence through future input dependence \cite{retro1,retro2,retro3,Wharton2} have recently been surveyed in \cite{Wharton} (Section VI) . We will here not go through these toy models again, but instead briefly introduce  existing approaches to an underlying theory that give rise to Superdeterminism. 

These approaches, needless to say, are still in their infancy. They leave open many questions and it might well turn out that none of them is the right answer. We do believe, however, that they present a first step on the way towards a satisfactory solution of the measurement problem. 

\subsection{Invariant Set Theory}
\label{IST}

Invariant Set Theory (IST) \cite{Palmer:2009mxd, Palmer:2018mxd} arose from an earlier realisation \cite{Palmer:1995mxd} that, suitably formulated, nonlinear dynamics could provide the basis for a deterministic theory of quantum physics which was not counterfactually complete and therefore could violate Statistical Independence thus avoiding nonlocality. More specifically, IST is a deterministic theory based on the assumption that the laws of physics at their most primitive derive from the geometry of a fractal set of trajectories, or histories, $I_U$, in state space. States of physical reality -- the space-time that comprises our universe and the processes which occur in space-time -- are those and only those belonging to $I_U$; other states in the Euclidean space in which $I_U$ is  embedded, do not correspond to states of physical reality. Dynamical evolution maps points on $I_U$ to other points on $I_U$, whence $I_U$ is invariant under dynamical laws of evolution. In this theory, the basic element of $I_U$ is a fractal helix (in the sense that each trajectory in the helix, like a strand of rope, is itself a helix of finer-scale trajectories).

The link to quantum mechanics is made through the statistical properties of the helices which can be represented by complex Hilbert vectors and tensor products, where squared amplitudes and complex phases of the Hilbert vectors are necessarily described by rational numbers. 

IST provides some possible understanding of a key difference between the Liouville equation and the von Neumann-Dirac equation: the factor $i \hbar$. Planck's constant has the dimension of state space (momentum times position) and hence provides an inherent size to any geometric structure in state space, such as $I_U$. This inherent size is provided by the radius of a helix of $I_U$. Based on the $U(1) \sim SO(2)$ isomorphism, the square root of minus one is consistent with a rotational symmetry of the helical nature of the trajectories of $I_U$. 

Since it is formulated in terms of complex Hilbert states, IST violates Bell inequalities exactly as does quantum theory. It does this not only because Statistical Independence is violated (the fractal gaps in $I_U$ correspond to states of the world associated with certain counterfactual measurement settings, which by construction are not ontic), it also violates the Factorisation assumption of Bell's theorem and hence is Bell-nonlocal. Because the set of helices has fractal structure, the $p$-adic metric, rather than Euclidean metric is a natural measure of distance in state space.  
 
Importantly, violation of Statistical Independence and Factorisation only occur when one considers points which do not lie on  $I_U$. From a Hilbert state perspective, they are associated with Hilbert States where either squared amplitudes or complex phases of Hilbert States cannot be described by rational numbers. Hence, the violations of Statistical Independence and Factorisation in IST arise because certain putative counterfactual states are mathematically undefined; without these violations there would be mathematical inconsistency. Importantly, such counterfactual states do not correspond to physically possible processes in space-time. If Statistical Independence and Factorisation are weakened to only allow processes which are expressible in space time and hence are physically possible (`Statistical Independence on $I_U$' and `Factorisation on $I_U$'), then IST is consistent with both free choice and locality. 

In IST, the measurement process is described by state-space trajectories that cluster together near detector eigenstates. In this approach, the measurement problem has been largely nullified because the statistical state space of the trajectory segments that lead to those detector eigenstates is no longer the whole Hilbert space, but instead the space whose elements have finite squared amplitudes and complex phases. In this sense, IST does not `complete' quantum theory. Rather, it is a replacement for quantum theory, even at the pre-measurement unitary stage of evolution.

The fractal attractor which defines $I_U$ can be considered a future asymptotic property of some more classical like governing differential equations of motion: start from any point in state space and the trajectory will converge onto it only as $t \rightarrow \infty$. The invariant set is therefore operationally incomputable in much the same way that the event horizon of a black hole is. 

\subsection{Cellular Automata}
\label{thooft}

The Cellular Automata approach to Superdeterminism \cite{Hooft:2014kka} is a model that employs a time-evolution which proceeds in discrete time-steps on a grid. It uses a language similar to quantum mechanics, in that the state-space is spanned by vectors in a Hilbert-space. These vectors can, as usual, be brought into superpositions. However, it is then postulated that states which result in superpositions that we do not observe are not ontic. It follows from this that an initial state which gave rise to an unobserved outcome was not ontic either.
A variety of simple toy models have been discussed in \cite{Hooft:2014kka}. 

In this approach there is strictly speaking only one ontic state in the theory, which is the state that the universe is in. The requirement that the final state must correspond to the classical reality which we observe induces constraints at earlier times. These constraints give rise to nonlocal correlations which result in a violation of Statistical Independence. 

The challenge for this approach is to render this theory predictive. As was noted in \cite{Hooft:2014kka}, selecting the ontological state requires a measure for the initial states of the universe:

\begin{quote}
``Bell’s theorem
requires more hidden assumptions than usually thought: {\sl The quantum theory only
contradicts the classical one if we assume that the ‘counterfactual modification’
does not violate the laws of thermodynamics.} In our models, we must assume that it does.'' (emphasis original)
\end{quote}

\noindent It is presently unclear from where such a thermodynamic-like measure comes. 

\subsection{Future-bounded Path Integrals}
\label{path}

The path integral approach to Superdeterminism \cite{future} rests on the observation that the Feynman path integral has a future input dependence already, which is the upper time of the integration. However, in the usual path integral of quantum mechanics (and, likewise, of quantum field theory), one does not evaluate what is the optimal future state that the system can evolve into. Instead, one posits that all of the future states are realized, which results in a merely probabilistic prediction.

The idea is then to take a modified path integral for the combined system of detector and prepared state and posit that in the underlying theory the combined system evolves along merely one possible path in state space that optimizes a suitable, to-be-defined, function. This function must have the property that initial states which evolve into final states  containing superpositions of detector eigenstate states are disfavoured, in the sense that they do not optimize the function. Instead, the optimal path that the system will chose is one that ends up in states which are macroscopically classical. One gets back normal quantum mechanics by averaging over initial states of the detector.

This approach solves the measurement problem because the system does deterministically evolve into one particular measurement outcome. Exactly which outcome is determined by the degrees of freedom of the detector that serve as the ``hidden variables''. Since it is generically impossible to exactly know all the detector's degrees of freedom, quantum mechanics can only make probabilistic predictions.

The challenge of this approach is to find a suitable function that actually has this behaviour. 

\section{Experimental Test}

It is clear that the above discussed theoretical approaches to Superdeterminism require more work. However, such theories have general properties that, with some
mild assumptions, tell us what type of experiment has the potential to reveal deviations from quantum mechanics. 

To see this, we first note that typical experiments in the foundations of quantum mechanics probe physics at low energies, usually in the range of atomic physics. It is, however, difficult to come up with any model that equips known particles with new degrees of freedom accessible at such low energies. The reason is that such degrees of freedom would change the phase-space of standard model particles. Had they been accessible with any experiment done so far, we would have seen deviations from the predictions of the standard model, which has not happened. 

It is well possible to equip standard model particles with new degrees of freedom if those are only resolvable at high energies (examples abound). But in this case
the new degrees of freedom do not help us with solving the measurement problem exactly because we assumed that they do not play a role at the relevant energies. 

If one does not want to give up on this separation of scales, this leaves the possibility that the hidden variables are already known degrees of freedom of particles which do not comprise the prepared state. Moreover, they are only those degrees of freedom that are resolvable at the energy scales under consideration. 

The next thing we note is that all presently known deterministic, local theories have the property that states that were close together at an initial time will remain close for some while. In a superdeterministic theory, states with different measurement settings are distant in state-space, but changes to the hidden variables that do not also change the measurement setting merely result in different measurement outcomes and therefore correspond to states close to each other. 

Since the theory is deterministic, this tells us that if we manage to create a time-sequence of initial states similar to each other, then the measurement outcomes should also be similar. This means conceretely that rather than fulfilling the Born-rule, such an experiment would reveal time-correlations in the measurement outcomes. The easiest way to understand this is to keep in mind that if we were able to exactly reproduce the initial state, then in a superdeterministic theory the measurement outcome would have to be the same each time, in conflict with the predictions of quantum mechanics. 

This raises the question how similar the initial states have to be for this to be observable. Unfortunately, this is not a question which can be answered in generality; for this one would need a theory  to make the corresponding calculation. However, keeping in mind that the simplest case of hidden variables are the degrees of freedom of other particles and that the theory is local in the way we are used to it, the obvious thing to try is minimizing changes of the degrees of freedom of the detecting device. Of course one cannot entirely freeze a detector's degrees of freedom, for then it could no longer detect something. But one can at least try to prevent non-essential changes, i.e., reduce noise.

This means concretely that one should make measurements on states prepared as identically as possible with devices as small and cool as possible in time-increments as small as possible. 

This consideration does not change much if one believes the hidden variables are properties of the particle after all. In this case, however, the problem is that preparing almost identical initial states is impossible since we do not know how to reproduce the particle's hidden variables. One can then try to make repeated measurements of non-commuting observables on the same states, as previously laid out in \cite{Hossenfelder:2011ct}. 

The distinction between the predictions of quantum mechanics and the predictions of the underlying, superdeterministic theory is not unlike the distinction between climate predictions and weather forecasts. So far, with quantum mechanics, we have made predictions for long-term averages. But even though we are in both cases dealing with a non-linear and partly chaotic system, we can in addition also make short-term predictions, although with limited accuracy. The experiment proposed here amounts to recording short-term trends and examining the data for regularities that, according to quantum mechanics alone, should not exist. 

Needless to say, the obvious solution may not be the right one and testing Superdeterminism may be more complicated than that. But it seems reasonable to start with the simplest and most general possibility before turning to model-specific predictions. 

\section{Discussion}
\label{disc}

The reader may have noticed a running theme in our discussion of Superdeterminism, which is that objections raised against it are deeply rooted in intuition that is, ultimately, based on the classical physics we experience with our own senses.

\begin{wrapfigure}{r}{0.43\textwidth}
  \begin{center}
    \includegraphics[width=0.88\textwidth]{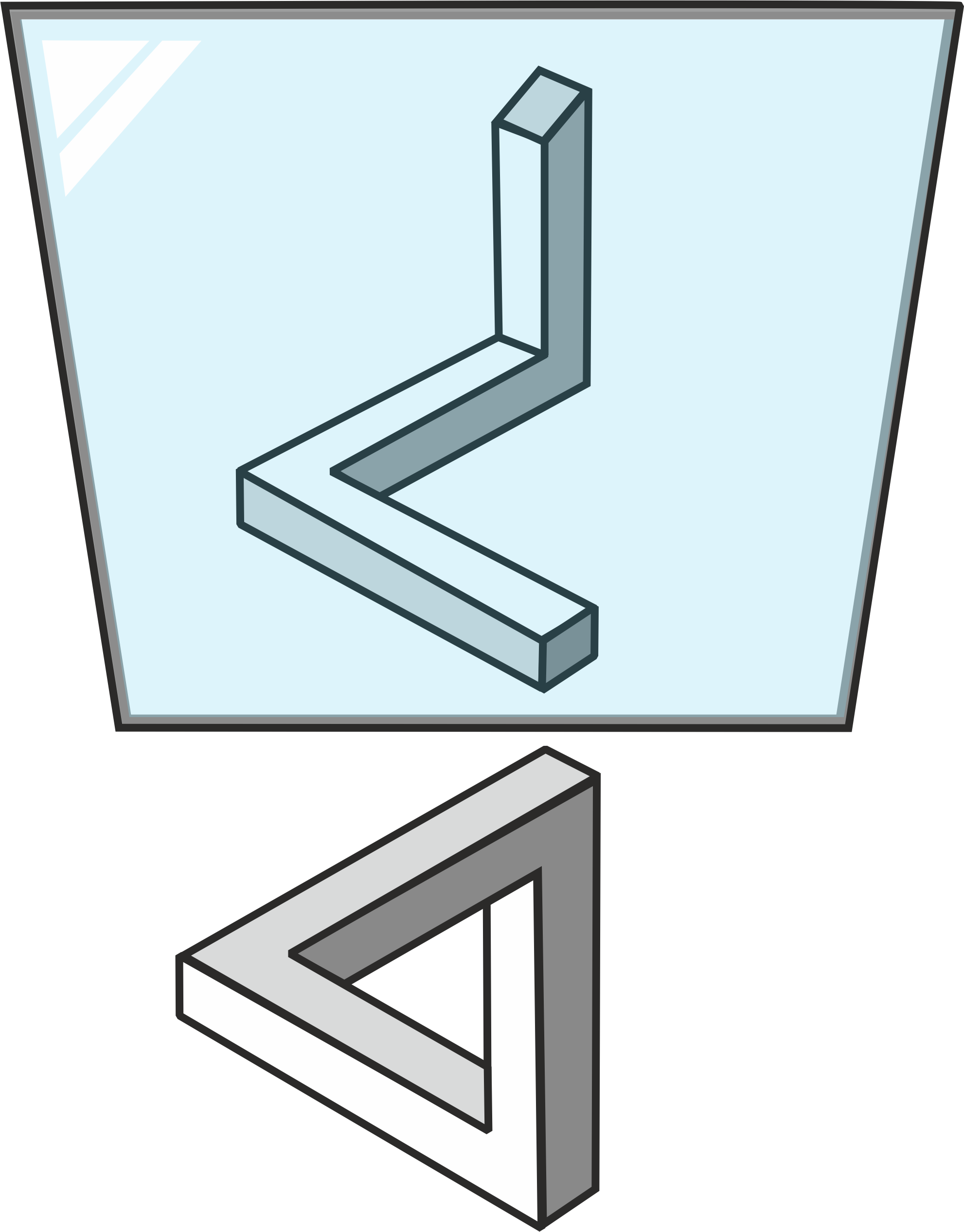}
  \end{center}
  {\caption{Penrose's ``impossible'' triangle, placed in front of a mirror, turns out to be not so impossible. \label{fig1} }}
\end{wrapfigure}

But these intuitions can mislead us. For an illustration, consider Penrose's impossible triangle (see Fig \ref{fig1}, bottom). If we see a two-dimensional drawing of the triangle, we implicitly assume that any two arms come closer as they approach a vertex. This raises the impression that the object is impossible to realize in 3-dimensional space. However, the supposedly impossible triangle can be built in reality. The object shown in Fig \ref{fig1}, top, seen from the right direction, reproduces what is shown in the 2-dimensional drawing. From any other direction, however, it becomes clear that our intuition has led us to improperly assume two arms necessarily become close as they approach a common vertex. 

We believe that the uneasiness we bring to considering Superdeterminism stems from a similar intuitive, but ultimately wrong, idea of closeness. In this case, however, we are not talking about closeness in position space but about closeness in the state-space of a theory.

Faced with trying to quantify the ``distance'' between two possible states of the universe our intuition is to assume that it can be measured in state space by the same Euclidean metric we use to measure distance in physical space. This indeed is the basis of Lewis's celebrated theory of causality by counterfactuals: of two possible counterfactual worlds, the one that resembles reality more closely is presumed closer to reality \cite{Lewis}. But is this really so? In number theory there is an alternative to the class of Euclidean metrics (and indeed according to Ostrowsky's theorem it is the only alternative): the $p$-adic metric \cite{Katok}. The $p$-adic metric is to fractal geometry as the Euclidean metric is to Euclidean geometry. The details do not need to concern us here, let us merely note that two points that are close according to the Euclidean metric may be far away according to the $p$-adic metric. 


This means from the perspective of the $p$-adic metric, the distance between the actual world where the parity of the millionth digit of the input to Bell's pseudo-random number generator was, say, 0, and the counterfactual world where the parity was a 1 could be very large, even though it is small using an Euclidean measure of distance. A theory that seems fine-tuned with respect to the latter metric would not be fine-tuned with respect to the former metric. Like with Penrose's triangle, the seemingly impossible becomes understandable if we are prepared to modify our intuition about distance. 

But our intention here was not merely to draw attention to how classical intuition may have prevented us from solving the measurement problem. Resolving the measurement problem with Superdeterminism may open the door to solving further problems in the foundations of physics. As has been previously noted \cite{Hossenfelder:2012uy}, our failure to find a consistent quantum theory of gravity may be due, not to our lacking understanding of gravity, but to our lacking understanding of quantization. The same problem may be behind some of the the puzzles raised by the cosmological constant. It is further a long-standing conjecture that dark matter is not a new type of particle but instead due to a modification of gravity. We know from observations that such a modification of gravity is parametrically linked to dark energy \cite{one}. The reasons for this connection are currently not well-understood, but completing quantum mechanics, or replacing it with a more fundamental theory, might well be the key to solving these problems.

Finally, let us point out that the technological applications of quantum theory become more numerous by the day. Should we discover that quantum theory is not fundamentally random, should we succeed in developing a theory that makes predictions beyond the probabilistic predictions of quantum mechanics, this would likely also result in technological breakthroughs. 

\section{Conclusion}

We have argued here that quantum mechanics is an incomplete theory and completing it, or replacing it with a more fundamental theory, will necessarily require us to accept violations of Statistical Independence, an assumption that is sometimes also, misleadingly, referred to as Free Choice. We have explained why objections to theories with this property, commonly known as superdeterministic, are ill-founded. 

Since the middle of the past century, progress in the foundations of physics has been driven by going to shorter and shorter distances, or higher and higher energies, respectively. But the next step forward might be in an entirely different direction, it might come from finding a theory that does not require us to hand-draw a line between microscopic and macroscopic reality.

\section*{Acknowledgements}

SH gratefully acknowledges support from the Franklin Fetzer Fund. TNP gratefully acknowledges support from the Royal Society.

\end{document}